# Towards Electrical-Current Control of Quantum States in Spin-Orbit-Coupled Matter

*Gang Cao*[*]
*Department of Physics, University of Colorado at Boulder, Boulder, Colorado 80309, USA*

Novel materials, which often exhibit surprising or even revolutionary physical properties, are necessary for critical advances in technologies. Simultaneous control of structural and physical properties via a small electrical current is of great significance both fundamentally and technologically. Recent studies demonstrate that a combination of strong spin-orbit interactions and a distorted crystal structure in magnetic Mott insulators is sufficient to attain this long-desired goal. In this *Topical Review*, we highlight underlying properties of this class of materials and present two representative antiferromagnetic Mott insulators, namely, *4d*-electron based $Ca_2RuO_4$ and *5d*-electron based $Sr_2IrO_4$, as model systems. In essence, a small, applied electrical current engages with the lattice, critically reducing structural distortions, which in turn readily suppresses the antiferromagnetic and insulating state and subsequently results in emergent new states. While details may vary in different materials, at the heart of these phenomena are current-reduced lattice distortions, which, via spin-orbit interactions, dictate physical properties. Electrical current, which joins magnetic field, electric field, pressure, light, etc. as a new external stimulus, provides a new, key dimension for materials research, and also pose a series of intriguing questions that may provide the impetus for advancing our understanding of spin-orbit-coupled matter. This *Topical Review* provides a brief introduction, a few hopefully informative examples and some general remarks. It is by no means an exhaustive report of the current state of studies on this topic.

---

[*] gang.cao@colorado.edu

# Table of Content

# I. Overview

Electrical-current control of quantum states has been a long-sought but nearly elusive goal of science and technology. Recent studies of spin-orbit-coupled materials, particularly, *4d-* and *5d-*transition metal oxides [1-4], provide strong evidence that this goal may be finally within reach [5-12].

*4d-* and *5d-*transition metal materials (*4d/5d* materials) feature strong spin-orbit interactions (SOI), extended *d*-electron orbitals, thus reduced (although still significant) on-site Coulomb interaction U, compared to *3d*-transition metal materials. As shown in **Table 1**, *4d/5d* materials host a unique hierarchy of energy scales defined

Table 1. Comparison between compounds with 3d, 4d, and 5d electrons.

| Electron Type | U(eV) | $\lambda_{so}$(eV) | Key Interactions |
|---|---|---|---|
| 3d | 5-7 | 0.01-0.1 | $U \gg \lambda_{so}$ |
| 4d | 0.5-3 | 0.1-0.3 | $U > \lambda_{so}$ |
| 5d | 0.4-2 | 0.1-1 | $U \sim \lambda_{so}$ |

U = Coulomb interaction; $\lambda_{SO}$ = SOI

by comparable and competing spin-orbit and Coulomb interactions. This energy setting generates a rare, delicate interplay between the fundamental interactions and leaves these materials precariously balanced on the border between different ground states, and extremely susceptible to even small, external stimuli. These stimuli strongly couple with the lattice, thus resulting in emergent novel quantum states [1, 4, 13, 14]. Electrical current, as a new external stimulus (joining magnetic field, pressure, light, etc.), is surprisingly effective in coupling with the lattice, thus quantum state via SOI in certain *4d/5d* materials, according to recent studies on this topic [5-12]. As schematically illustrated in **Fig.1**, applied current relaxes and expands the lattice, thus precipitating new electronic structures or ground states. It is not surprising that current control of quantum states has rapidly become a key research topic [5-12]. Remarkably, control via application of a small current is different from control via application of static or low-frequency electric fields, such as electric-field-controlled magnetic properties in multiferroics, which is an important and yet vastly different topic that has been extensively studied over the last two decades [15-16, for example].

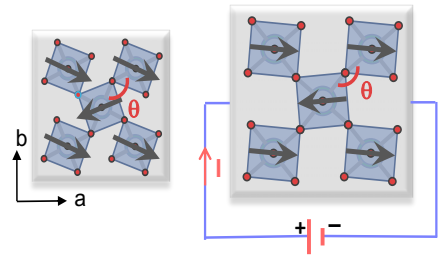

*Fig. 1. A schematic illustration of current-controlled structural and physical properties. The black arrows are magnetic moments strongly locked with the lattice. Applied current I engages with the lattice and relaxes lattice distortions via SOI.*

In this this *Tropical Review*, we first present a brief survey of relevant, underlying properties of *4d/5d* oxides (*Section II*), then focus on two representative materials recently studied (*Section III*) and finally conclude the article with some general remarks and an array of questions arising from the recent studies (*Section IV*). This article provides only a glimpse of this exciting, rapidly evolving research area, and is by no means an exhaustive report of the current state of experimental studies on the topic, which is still in its infancy.

## II. Fundamental Characteristics of 4d- and 5d-Transition Metal Oxides

### A. Strong electron-lattice coupling

One key characteristic of the *4d/5d*-electron transition elements is their more extended *d*-orbitals, compared to those of their *3d*-electron counterparts. Consequently, strong *p-d* hybridization and electron-lattice coupling, along with the reduced intraatomic Coulomb interaction U and Hund's rule coupling $J_H$, are expected in these systems (see **Table 1**). The deformations and relative orientations of corner-shared $MO_6$ (M=transition metal) octahedra determine the crystalline-electric-field level splitting and the electronic band structure, and hence the ground state. The physical properties of these materials are thus highly sensitive to lattice distortions and dimensionality and susceptible to external stimuli, such as application of magnetic field, pressure, light, chemical doping [1-4], or, as recently discovered, electrical current [5-12]. Some of these characteristics are well demonstrated by contrasting physical properties of the Ruddlesden-Popper (RP) series of $Ca_{n+1}Ru_nO_{3n+1}$ and $Sr_{n+1}Ru_nO_{3n+1}$ where n (=1, 2, 3, ∞) is the number of Ru-O layers per unit cell [17-30]. The ground states of this class of materials sensitively depend on the ionic radius of the alkaline earth cation, which is 1.00 Å and 1.18 Å for Ca and Sr, respectively, thus distortions and rotations/tilts of $RuO_6$ octahedra. As a result, the less structurally distorted $Sr_{n+1}Ru_nO_{3n+1}$ compounds are metallic and tend to be ferromagnetic (FM) (see **Figs. 2a** and **2b**), whereas the more structurally distorted $Ca_{n+1}Ru_nO_{3n+1}$ compounds are all proximate to a metal-insulator transition and prone to antiferromagnetic (AFM) order (see **Figs. 2c** and **2d**). Such a distinct characteristic is at the heart of numerous novel phases in the RP ruthenates uncovered via external stimuli that couple to the lattice [1].

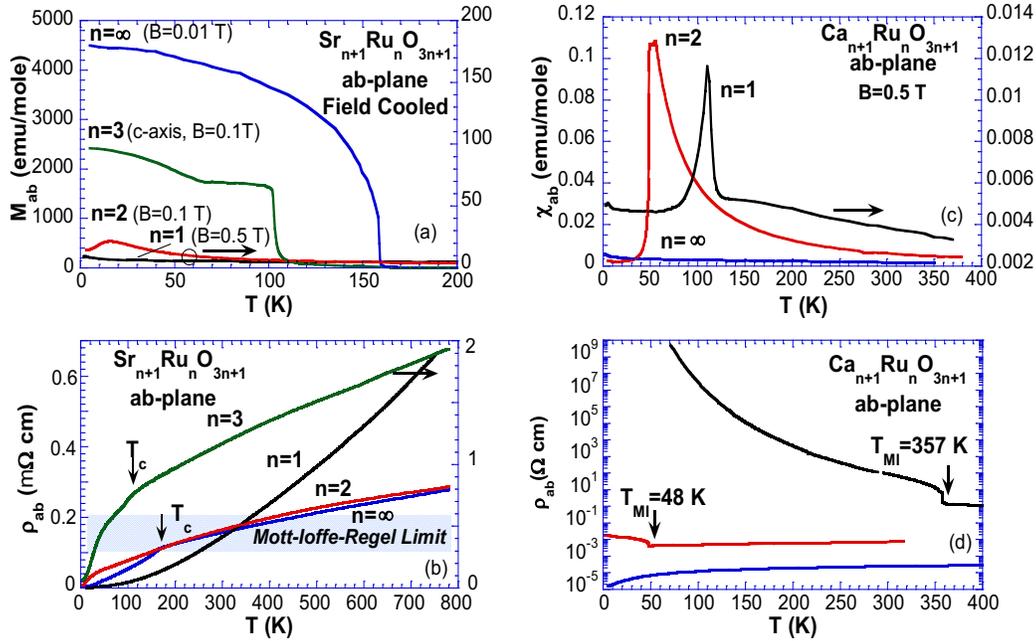

***Fig.2.*** *Magnetic susceptibility χ, magnetization M (upper panels) and resistivity ρ (lower panels) as a function of temperature for R-P series $Sr_{n+1}Ru_nO_{3n+1}$ (left column) and $Ca_{n+1}Ru_nO_{3n+1}$ (right column) with n=1,2,3 and ∞. Note the sharp differences in the ground state between the Ca- and Sr-compounds and n-dependence of M, χ and ρ [1].*

**B. Strong spin-orbit interactions**

The inherently strong SOI are another driver of the physics of *4d/5d* oxides (**Table 1**) [1-4]. The phenomenology of the SOI and their fundamental consequences for material properties have been justifiably neglected until recently, due to the pervasive emphasis placed upon the 3*d*-elements in the conduct of both basic research and technical development. Nevertheless, traditional arguments would suggest that *5d*-electron based oxides should be more metallic and less magnetic than oxides containing 3*d*, 4*f* or even *4d* elements, because *5d*-electron orbitals are more extended in space, which leads to increased electronic bandwidth. This conventional wisdom conflicts with early experimental observations in almost all existing *5d*-electron based iridates such as $Sr_{n+1}Ir_nO_{3n+1}$ [31-34] and $BaIrO_3$ [35], which are conversely both AFM and insulating (see **Fig. 3** and **Table 2**).

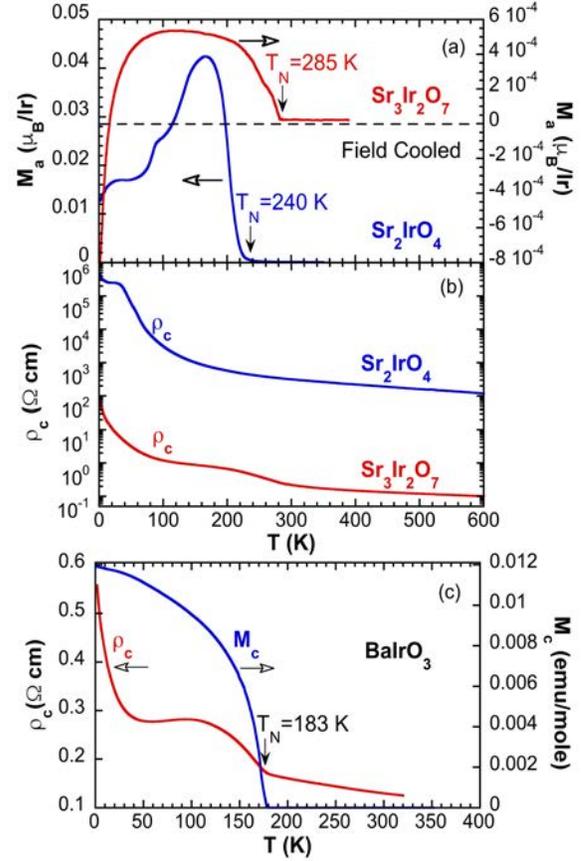

*Fig. 3. Temperature dependence of (a) the a-axis magnetization $M_a$, and (b) the resistivity $\rho$ for $Sr_2IrO_4$ (blue) and $Sr_3Ir_2O_7$ (red) and (c) the c-axis resistivity $\rho_c$ (red, left scale) and c-axis magnetization $M_c$ (blue, right scale) for $BaIrO_3$.*

It is now recognized that a critical underlying mechanism for these unanticipated states is that strong SOI that split the otherwise broad *5d*-band and vigorously competes with Coulomb interactions, crystalline electric fields, and Hund's rule coupling. The so-called "$J_{eff} = ½$" insulating state in $Sr_2IrO_4$ served as an early signal that the strong SOI in iridates might have unique consequences [37-39]. Since the SOI is a relativistic effect proportional to the atomic number Z $Sr_{n+1}Ir_nO_{3n+1}$ [31-34] and $BaIrO_3$ [35], which are conversely both AFM and insulating (see **Fig. 3** and **Table 2**).

6The inherently strong SOI are another driver of the physics of *4d/5d* oxides (**Table 1**) [1-4]. The phenomenology of the SOI and their fundamental consequences for material properties have been justifiably neglected until recently, due to the pervasive emphasis placed upon the 3*d*-elements in the conduct of both basic research and technical development. Nevertheless, traditional arguments would suggest that *5d*-electron based oxides should be more metallic and less magnetic than oxides containing 3*d*, 4*f* or even *4d* elements, because *5d*-electron orbitals are more extended in space, which leads to increased electronic bandwidth. This conventional wisdom conflicts with early experimental observations in almost all existing *5d*-electron based iridates such as

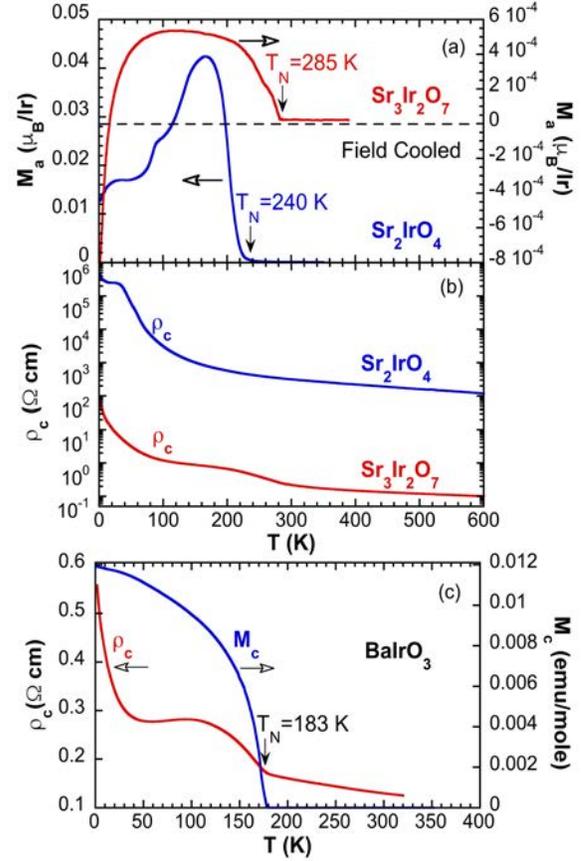

*Fig. 3. Temperature dependence of (a) the a-axis magnetization $M_a$, and (b) the resistivity $\rho$ for $Sr_2IrO_4$ (blue) and $Sr_3Ir_2O_7$ (red) and (c) the c-axis resistivity $\rho_c$ (red, left scale) and c-axis magnetization $M_c$ (blue, right scale) for $BaIrO_3$.*

It is now recognized that a critical underlying mechanism for these unanticipated states is that strong SOI that split the otherwise broad *5d*-band and vigorously competes with Coulomb interactions, crystalline electric fields, and Hund's rule coupling. The so-called "$J_{eff} = ½$" insulating state in $Sr_2IrO_4$ served as an early signal that the strong SOI in iridates might have unique consequences [37-39]. Since the SOI is a relativistic effect proportional to the atomic number Z or $Z^2$ [40, 41], it has an approximate strength of 0.4 eV in the iridates (compared to around 20 meV in 3d materials), and splits the $t_{2g}$ 5d-bands into states with $J_{eff}$ = 1/2 and $J_{eff}$ = 3/2, the latter having lower energy [37, 38] (see **Table 1**). Ir$^{4+}$ ($5d^5$) ions provide five 5d-electrons to bonding states, four of them fill the lower $J_{eff}$ = 3/2 bands, and one is left to partially fill the $J_{eff}$ = 1/2 band in which the Fermi level $E_F$ resides. The $J_{eff}$ = 1/2 band is so narrow that even a reduced on-site Coulomb repulsion (U ~ 0.5 eV, due to the extended nature of 5d-electron orbitals) is sufficient to open a small gap Δ that stabilizes an insulating state, as shown in **Fig. 4** [4, 37].

**Table 2. Examples of Insulating and Antiferromagnetic Iridates**

| System | Néel Temperature (K) | Ground State |
|---|---|---|
| $Sr_2IrO_4$ (n = 1) | 240 | Canted AFM insulator |
| $Sr_3Ir_2O_7$ (n = 2) | 285 | AFM insulator |
| $BaIrO_3$ | 183 | Canted AFM insulator |

SOI can be modified by electronic correlations in the same way as the interatomic Coulomb interactions can be screened by itinerant electrons. It has been suggested that the effect of the Coulomb correlations can actually enhance the SOI in 4d electron systems such as $Sr_2RhO_4$ [42].

It is worth mentioning that $Sr_2RhO_4$ is similar to $Sr_2RuO_4$ and $Sr_2IrO_4$ both electronically and structurally, but its ground state is fundamentally different from those of the other two compounds. $Sr_2RhO_4$ hosts a Rh$^{4+}$ ion with five 4d-electrons (compared to four 4d-electrons of the Ru$^{4+}$ ion in $Sr_2RuO_4$). It shares a crystal structure remarkably similar to that of $Sr_2IrO_4$; in particular, the RhO$_6$ octahedron rotates about the c axis by 10° [43]; this value is zero for $Sr_2RuO_4$ and 12° for $Sr_2IrO_4$, as discussed below. It is argued that this octahedral rotation facilitates a correlation-induced enhancement of the SOI by about 20%, that is, the SOI increases from the bare value of 0.16 eV to 0.19 eV [42]. Despite its similarities to the insulating $Sr_2IrO_4$, $Sr_2RhO_4$ is a paramagnetic metal instead because the SOI is still not strong enough to conspire with the Coulomb interaction to open an energy gap [44], as the case in $Sr_2IrO_4$ discussed above and shown in **Fig.4**. On the other hand, $Sr_2RhO_4$ is indeed in close proximity to an insulating state because of the octahedral rotation and the enhanced SOI -- with slight Ir doping for Rh, it becomes an insulator [45]. The less robust metallic state is also because the $t_{2g}$ bands near the Fermi surface are less dispersive in $Sr_2RhO_4$ than in $Sr_2RuO_4$, therefore more susceptible to the SOI-induced band shifts near the Fermi surface than in $Sr_2RuO_4$ [46]. This is in part because the $Rh^{4+}(4d^5)$ ion has five 4d electrons instead of four as in the $Ru^{4+}(4d^4)$ ion. Nevertheless, all this further highlights the very importance of both the electron-lattice coupling and SOI in 4d/5d materials.

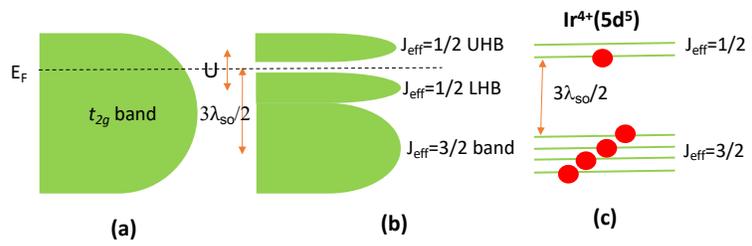

**Fig.4.** Band schematic: **(a)** The traditionally anticipated broad $t_{2g}$ band for 5d-electrons; **(b)** The splitting of the $t_{2g}$ band into $J_{eff}$=1/2 and $J_{eff}$=3/2 bands due to SOI; **(c)** $Ir^{4+}(5d^5)$ ions provide five 5d-electrons, four of them fill the lower $J_{eff}$ = 3/2 bands, and one electron partially fills the $J_{eff}$ = 1/2 band where the Fermi level $E_F$ resides.

A great deal of theoretical work has appeared in response to early experiments on iridates and motivated enormous activity in search of novel states in these materials. It is intriguing that many proposals have met limited experimental confirmation thus far. A good example is $Sr_2IrO_4$, an extensively studied spin-orbit-coupled material [4]. It is widely anticipated that with slight electron doping, $Sr_2IrO_4$ should be a novel superconductor because of its apparent similarities to those of $La_2CuO_4$ [47]. However, there has been no experimental confirmation of superconductivity characterized by zero resistivity and the Meissner effect, despite many years of experimental efforts. It is now recognized that the absence of the predicted superconductivity is due primarily to inherently severe structural distortions that suppress superconductivity [4, 13, 48]. A recent study reveals that the insulating state in $Sr_2IrO_4$ persists at megabar pressures, once again highlighting an overwhelming effect of structural distortions that prevent the expected onset of metallization, despite significant band broadening at 185 GPa [49]. On the other hand, slightly weakening the structural distortions via a newly developed field-editing technology during crystal growth nearly diminishes the insulating state in $Sr_2IrO_4$, according to a recent study [50]. In short, the lack of experimental confirmation of theoretical predictions of new states underscores a critical role of subtle structural distortions that dictate the low-energy Hamiltonian.

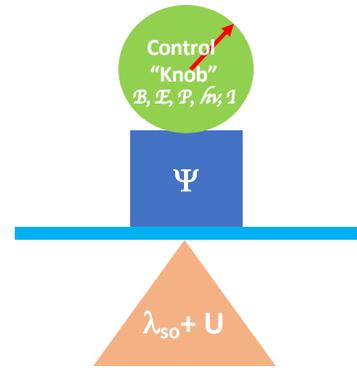

**Fig.5.** A schematic of the high susceptibility of a quantum state $\Psi$ supported by $\lambda_{so}$ and U to external stimuli, such as magnetic field B, electrical field E, pressure P, light hv or electrical current I.

However, it is precisely this unique characteristic that permits small external stimuli, such as applied current, to readily engage with the lattice, and, via strong magnetoelastic coupling, control magnetic and electronic states in materials hosting *a delicate interplay between $\lambda_{so}$ and U*, as schematically illustrated in **Fig.5**.

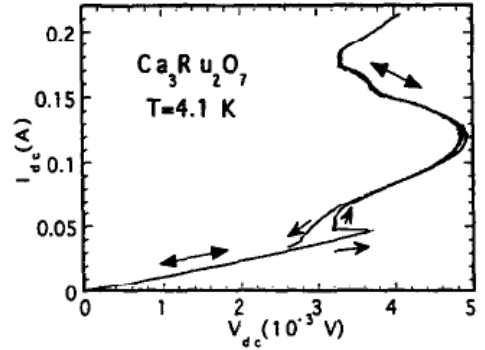

*Fig. 6. $Ca_3Ru_2O_7$:* Current-voltage characteristics: S-shaped NDR [51].

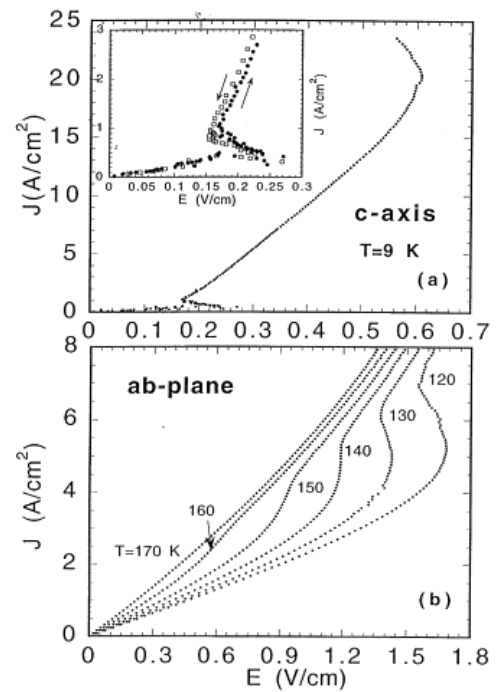

*Fig. 7. $BaIrO_3$:* S-shaped NDR for current along (a) the c-axis and (b) the ab-plane. The inset shows details of the noisy I–V characteristics at low current and the ohmic behavior for I =2 mA [36].

### III. Electrical-Current Control of Quantum States

In this section, we first review early observations of non-Ohmic I-V characteristics in some *4d* and *5d* transition metal oxides, which serve as an early sign of current-controlled behavior, and then present and discuss two model systems, namely, Sr$_2$IrO$_4$ and Ca$_2$RuO$_4$, in which application of a small current causes drastic changes in both structural and physical properties and emergent novel states otherwise unattainable.

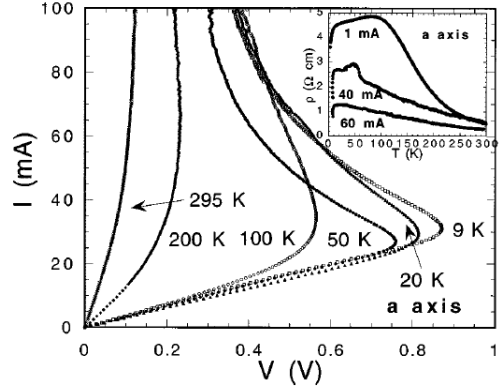

*Fig.8. Sr$_2$IrO$_4$: I-V characteristics for various temperatures. Inset: ρ along the a-axis vs temperature for various currents [34].*

It is important to be pointed out that Joule heating could cause spurious behavior. Therefore, extreme care must be taken to ensure that current-controlled phenomena are intrinsic. Results presented in this *Topical Review* convincingly eliminate a role of Joule heating [6,8]. This issue is briefly discussed in *Sections III B-C* below. Nevertheless, studies of current-controlled phenomena require robust, innovated techniques that must allow adequate measurements of samples with effective control of Joule heating and temperature.

### A. Early experimental observations of current-controlled phenomena in 4d/5d oxides

Early studies already suggested that electronic properties of certain *4d/5d* oxides are sensitive to applied electrical current. In the late 1990's, it was found that Ca$_3$Ru$_2$O$_7$ [51], BaIrO$_3$ [36] and Sr$_2$IrO$_4$ [34] exhibit a "S"-shaped, negative differential resistivity (NDR) (see **Figs. 6-8**). The NDR is in general attributed to either an "electro-thermal" effect or a "transferred carrier effect" in which a current promotes carriers from a high- to a low-mobility band, as in the Gunn effect. The more common form of NDR is manifest in "N"-shaped characteristics [52-56]. Alternatively, an "S"-shaped NDR has been observed in memory devices and a few bulk materials such as VO$_2$, CuIr$_2$S$_{4-x}$Se$_x$, and 1T-TaS$_2$ [52-58]. All of these bulk materials are characterized by a first-order metal-insulator transition without an AFM state. It is therefore peculiar for AFM Sr$_2$IrO$_4$ and BaIrO$_3$ to show the S-shaped NDR because these materials show no first-order metal-insulator transition (**Fig.3**) [4]. Furthermore, an early study also reveals a current-induced metallic state in $Sr_2IrO_4$ (Inset in **Fig. 8**) [34]. The current-reduced resistivity in $Ca_2RuO_4$ is also reported in a more recent study [59]. All results of these previous studies signaled an early sign that a combined effect of SOI and Coulomb interactions driving *4d/5d* oxides may result in an unusually high susceptibility to application of electrical current. This realization helped motivate more extensive studies on this topic in recent years. Indeed, discoveries of current-induced diamagnetism in $Ca_2RuO_4$ [5], simultaneous current-control of structural and physical properties in $Sr_2IrO_4$ [6] and nonequilibrium orbital states in $Ca_2RuO_4$ [7-12] have been reported, respectively, since 2017. These new discoveries have arguably inaugurated the new research topic on investigations of current-controlled materials.

### B. Model System: Spin-Orbit-Coupled Mott Insulator $Sr_2IrO_4$

The AFM insulator $Sr_2IrO_4$ has been known for more than two decades [31-34] but it is only in recent years that the AFM insulating state is recognized as a consequence of a combined effect of strong SOI and

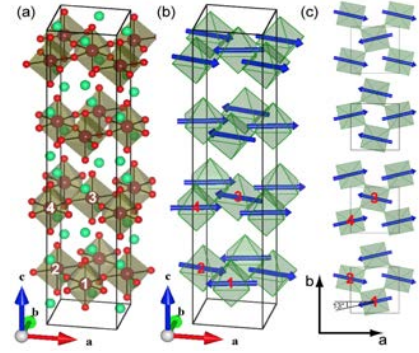

**Fig.9**. *$Sr_2IrO_4$: (a) Crystal structure. The $IrO_6$ octahedron rotates 11.8° about the **c** axis. The Ir atoms of the non-primitive basis are labeled 1, 2, 3, and 4. (b) The refined magnetic structure from single-crystal neutron diffraction measurements. (c) The same magnetic moment configuration projected on the basal planes [62]. Note that the magnetic canting closely tracks the $IrO_6$ rotation.*

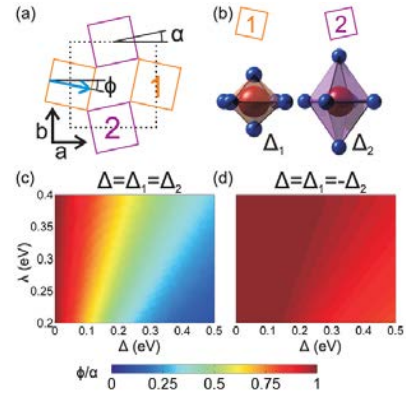

**Fig. 10.** *$Sr_2IrO_4$: (a) Illustration of an $IrO_2$ plane. The oxygen octahedra rotate about the **c**-axis by a creating a two-sublattice structure. The magnetic moments couple to the lattice and exhibit canting angles ϕ. (b) An unequal tetragonal distortion (Δ$_1$ and Δ$_2$) on the two sublattices as required by the $I4_1/a$ space group. (c) The ratio ϕ/α as a function of both SOI λ and Δ calculated for the case of uniform and (d) staggered (Δ$_1$ = − Δ$_2$) tetragonal distortion assuming U = 2.4 eV, Hund's coupling $J_H$ = 0.3 eV, hopping t = 0.13 eV, and α = 11.5° [64].*

Coulomb interactions [37] (see **Fig. 4**). As discussed above, this spin-orbit-coupled Mott insulator has an AFM transition at $T_N$=240 K and an electronic energy gap $\Delta \leq 0.62$ eV [38, 60, 61]. It adopts a tetragonal structure with $a = b = 5.4846$ Å and $c = 25.804$ Å with space-group $I4_1/acd$ (No. 142) [62,63], which is reduced to $I4_1/a$ (No. 88), according to more recent studies [64].

Two signature characteristics essential to current-controlled behavior are: **(1)** Rotation of the $IrO_6$-octahedra about the *c*-axis by approximately 12°, which corresponds to a distorted in-plane Ir1-O2-Ir1 bond angle θ and has a critical effect on the ground state; **(2)** The magnetic structure that features canted moments (0.208(3) $\mu_B$/Ir) within the basal plane [62]. This 13(1)°-canting of the moments away from the *a*-axis closely tracks the staggered rotation of the $IrO_6$ octahedra [63, 64] (see **Figs.9** and **10**), suggesting a strong interlocking of the magnetic moments to the lattice, which is absent in *3d* oxides [65].

The relationship between the $IrO_6$ rotation and magnetic canting in the iridate was first discussed in Ref.[66], in which a theoretical model proposed a strong magnetoelastic coupling in $Sr_2IrO_4$, and a close association between the magnetic canting and the ratio of the lattice parameter of the *c*-axis to the *a*-axis, as a result of the strong SOI. Such a strong locking of the moment canting to the $IrO_6$ -rotation (by 11.8(1)°) is experimentally manifest in later studies of neutron diffraction [62], X-ray resonant scattering [63] and second-harmonic generation (SHG) [64]. The SHG study also indicates that the *I41/a* space group requires a staggering of the sign of the tetragonal distortion, which helps explain the magnetoelastic locking, as illustrated in **Fig. 10** [64].

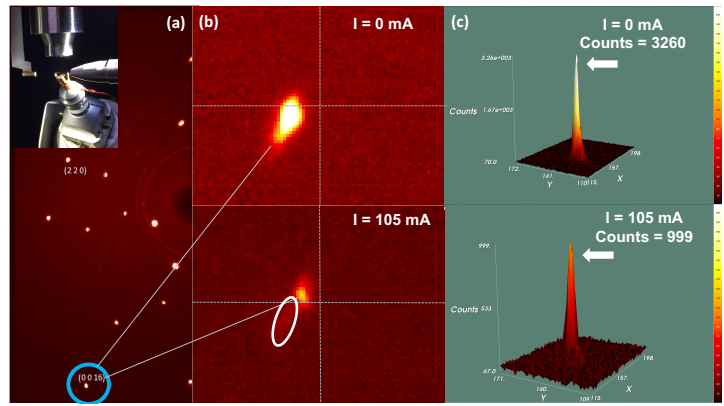

*Fig. 11. $Sr_2IrO_4$: Single-crystal x-ray diffraction with current I applied within the basal plane of the crystal. (a) Representative x-ray diffraction pattern of single crystal $Sr_2IrO_4$. The circled Bragg peak is (0,0,16). **Inset**: Sample mounting showing electrical leads and cryogenic gas feed. (b) Contrasting the (0016) peak location for I=0 mA (upper panel) and I=105 mA (lower panel). Note that the white oval outline marks the peak location for I=0 mA for comparison. (c) The intensity of the (0016) peak is 3260 for I = 0 mA (upper panel) and 999 for I = 105 mA (lower panel) [6].*

As demonstrated below, the strong interlocking between the lattice and magnetic moments is the key for the current control of quantum states in $Sr_2IrO_4$.

### 1. Current-controlled structural properties

Studies of crystal structures as a function of applied electrical current have seldom or never been reported before, in part because a dependence of crystal structure on electrical current was not at all conventionally anticipated. Our study on current-controlled structural properties [6] was initiated by our early experimental observations. Among them, single-crystal $Sr_2IrO_4$, which was placed under a high-power microscope aided with polarizing lights, showed a slight but visible color change when current was applied to it. It is this glimmer of the unusual behavior that has eventually led to comprehensive studies of this iridate [6], and later more *4d/5d* oxides [8], as functions of both electrical current and temperature.

For $Sr_2IrO_4$, **Fig. 11** exhibits changes of representative Bragg peaks at T = 200 K with applied current within the basal plane, and one of them is the peak with Miller index (0016) (**Fig 11a**). A close examination of this peak reveals that both the position and intensity of the (0016) peak undergo remarkable changes at the applied current I of 105 mA (see **Fig. 11b** and **Fig. 11c**). This peak shifts up and to the right with a threefold reduction in intensity from 3260 counts at I =0 mA to 999 counts at I = 105 mA (see **Fig. 11c**), suggesting significant shifts in the atomic positions. Similar changes are seen in other Bragg peaks [6].

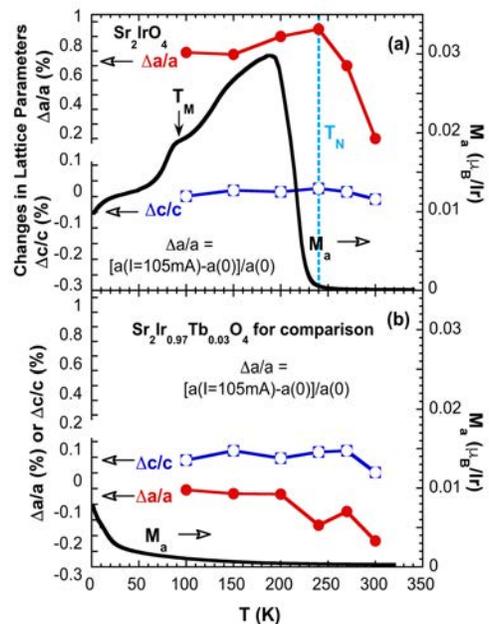

*Fig. 12. Current-controlled lattice change:* Current-controlled changes $\Delta a/a$ and $\Delta c/c$ and the a-axis magnetization $M_a$ (right scale) for **(a)** AFM $Sr_2IrO_4$ and **(b)** Isostructural, Paramagnetic $Sr_2Ir_{0.97}Tb_{0.03}O_4$ for comparison to **(a)**. Note that the scales for $\Delta a/a$, $\Delta c/c$ and $M_a$ are the same in **(a)** and **(b)** for contrast and comparison [6].

The above results have led to more detailed investigations of the crystal structure as functions of both current and temperature, which reveal an unexpectedly large lattice expansion due to applied current. In particular, at I=105 mA the *a*-axis elongates by nearly 1% ($\Delta a/a \equiv [a(I)-a(0)]/a(0)=1\%$) near $T_N$ = 240 K (see red curve in **Fig.12a**). In contrast, the *c*-axis changes only very slightly ($\Delta c/c < 0.1\%$) at the same current. The contrasting response of the *a*- and *c*-axis to current I indicates an important role of the basal-plane magnetic moments (**Fig.9**), implying the power of strong interlocking of cooperative magnetic order to the lattice (**Figs. 9-10**) [13, 66-68]. Indeed, the temperature dependence of $\Delta a/a$ closely tracks that of the *a*-axis magnetization, $M_a$, (black curve in **Fig. 12a**) whereas $\Delta c/c$ is essentially temperature independent.

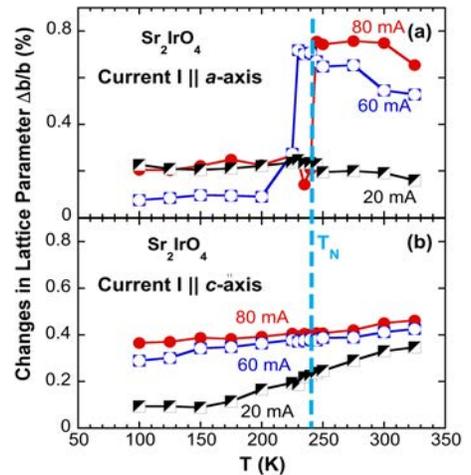

*Fig.13. $Sr_2IrO_4$: Anisotropic response of the lattice parameter **b** axis to current I applied along **(a)** the a-axis and **(b)** the c axis. Note that the scale for $\Delta b/b$ is the same for both **(a)** and **(b)** to facilitate comparisons. The effect of current applied along the a-axis is twice stronger than that of current applied along the c-axis. The abrupt expansion of the b-axis near $T_N$ observed in **(a)** further underscores the current effect and eliminates a role of heating effect.*

This is further confirmed by a controlled study of isostructural, paramagnetic $Sr_2Ir_{0.97}Tb_{0.03}O_4$, in which a 3% substitution of $Tb^{4+}$ for $Ir^{4+}$ leads to a disappearance of $T_N$ but conveniently preserves the original crystal structure and the insulating state [69] (In fact, that 3% Tb doping completely suppresses the AFM reaffirms the high sensitivity of the magnetic properties to slight lattice changes, but an energy level mismatch for the Ir and Tb sites weakens charge carrier hopping and causes a persistent insulating state [69]). This study indicates that changes in the lattice parameters or absolute values of $\Delta a/a$ and $\Delta c/c$ at I = 105 mA are very small (< 0.2%) and essentially temperature-dependent (**Fig. 12b**), The sharp contrast between **Fig. 12a**. and **12b** clearly points out a crucial role of long-range AFM in the current-induced lattice expansion [6]. Without application of current, the $a$ axis expands by no more than ~ 0.1% from 90 K to 300 K due to conventional thermal expansion [6], comparable to those of many materials [70].

The contrasting lattice expansion due to application of current (~1%) and temperature (~ 0.1%) once again indicates an effective coupling between current and the AFM state [6].

It is also found that the lattice parameters respond differently to current I applied to the $a$-axis and $c$-axis. As shown in **Fig.13**, the $b$-axis expands up to 0.8% near $T_N$ for I applied along the $a$-axis (**Fig.13a**) but only a half of that value for I applied along the $c$-axis (**Fig.13b**). Moreover, the abrupt jump in $\Delta b/b$ near $T_N$ tracks the magnetization M(T), further underscoring the interlocking of the canted moments to the lattice when I is along the *a* axis (**Fig.13a**). In contrast, *Δb/b* for I along the *c*-axis shows no similar behavior, suggesting a much weaker coupling of current and the moments (**Fig.13b**).

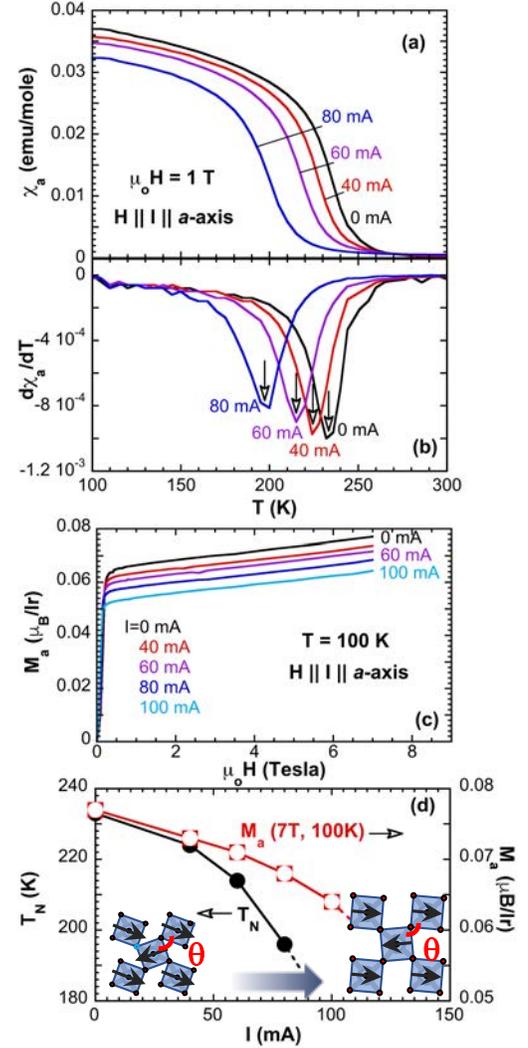

*Fig. 14. Sr$_2$IrO$_4$: Temperature dependence of (a) a-axis magnetic susceptibility $\chi_a(T)$ at a few representative currents, and (b) $d\chi_a(T)/dT$ clarifying the decrease in $T_N$ with I. (c) $M_a(H)$ at 100 K for a few representative currents. (d) Current dependence of $T_N$ and $M_a$. Diagrams schematically illustrate the current-controlled lattice expansion, $\theta$ (red) and Ir moments (black arrows) with I [6].*

The anisotropic response also rules out any effect of Joule heating. Should Joule heating play an important role, then its effect would be uniform or isotropic, rather than anisotropic as seen in **Fig.13**; additionally, the heating effect would be significantly stronger when current is applied along the *c*-axis because the *c*-axis resistivity is at least two orders of magnitude greater than the *a*-axis resistivity [4]. All this emphasizes that Joule heating does not play a significant role in the current-controlled phenomena in the iridate.

### 2. Current-controlled magnetic properties

Because of the strong coupling between the lattice and magnetic moments [62-66], it is compelling that the current-induced lattice expansion must cause changes in magnetic properties. As shown in **Fig. 14**, both the *a*-axis magnetic susceptibility $\chi_a(T)$ and the *a*-axis magnetization $M_a$ strongly respond to the current applied along the *a*-axis -- The AFM transition $T_N$ is suppressed by 40 K at I = 80 mA (**Figs. 14a** and **14b**) and the isothermal magnetization $M_a$ is reduced by 16% (**Fig.14c**). Magnetic canting is described by the Dzyaloshinsky-Moriya interaction, i.e., **D**·(**S**$_i$×**S**$_j$); the vector **D**, which measures distortions, approaches zero when the neighboring spins **S**$_i$ and **S**$_j$ become collinearly aligned. The magnetic changes presented in **Fig.14** are therefore understandable because the applied current relaxes the Ir-O-Ir bond angle θ, thus weakens magnetic canting and the overall AFM state, as schematically illustrated in **Fig.14d**.

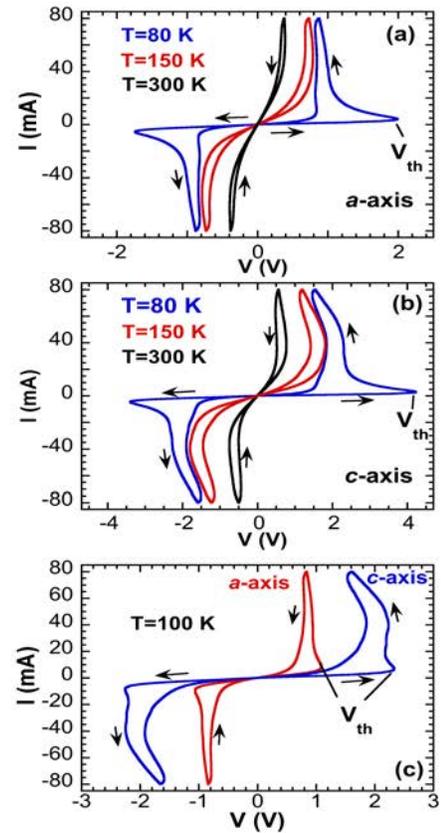

*Fig.15. Sr$_2$IrO$_4$: I-V curves at representative temperatures for: **(a)** Current applied along the **a**-axis, **(b)** along the **c**-axis, **(c)** both the **a**- and **c**-axis at T = 100 K. Arrows show the evolution of the current sweeps in **(a)** to **(c)**[6].*

### 3. *Current-controlled transport properties and non-Ohmic I-V characteristics*

The current-induced lattice expansion also enhances electron mobility in general and precipitates an unusual quantum switching effect in particular. As shown in **Figs. 15a-15c**, a linear I-V response during an initial current increase is followed by a sharp threshold voltage $V_{th}$, indicating a switching point where V abruptly drops with increasing I. This switching point is followed by another broad turning point that emerges at a higher current. A strong anisotropy in the IV-characteristics for the *a*-axis (**Fig.15a)** and *c*-axis (**Fig.15b)** is illustrated in **Fig.15c** [6].

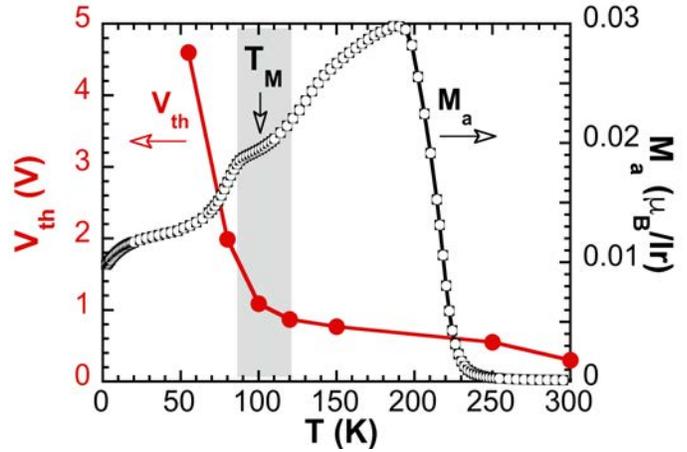

***Fig.16. $Sr_2IrO_4$: Correlation between the switching effect and magnetization*** *Temperature dependence of the threshold voltage $V_{th}$ for the **a**-axis (red) and **a**-axis magnetization $M_a(T)$ (black). Note the slope change of $V_{th}$ near $T_M$.*

Interestingly, the threshold voltage $V_{th}$ as a function of temperature shows a distinct slope change near a magnetic anomaly $T_M \approx 100$ K [13] (see **Fig. 16**). Early studies [13, 72, 73] have demonstrated that the magnetization $M_a$ undergoes additional anomalies at $T_M \approx 100$ K and 25 K (**Fig. 16,** black curve**,** right scale), due chiefly to moment reorientations [73]. This magnetic reorientation separates the different regimes of I-V behavior below and above $T_M \approx 100$ K (**Fig.16**). The concurrent changes in both $V_{th}$ and $M_a$ strongly indicates a close correlation between the I-V characteristics and magnetic state, and, more generally, a mechanism that is fundamentally different from that operating in other materials [53].

Furthermore, an emergent metallic state similar to that previously observed (inset in **Fig. 8**) also occurs at 20 mA [6].

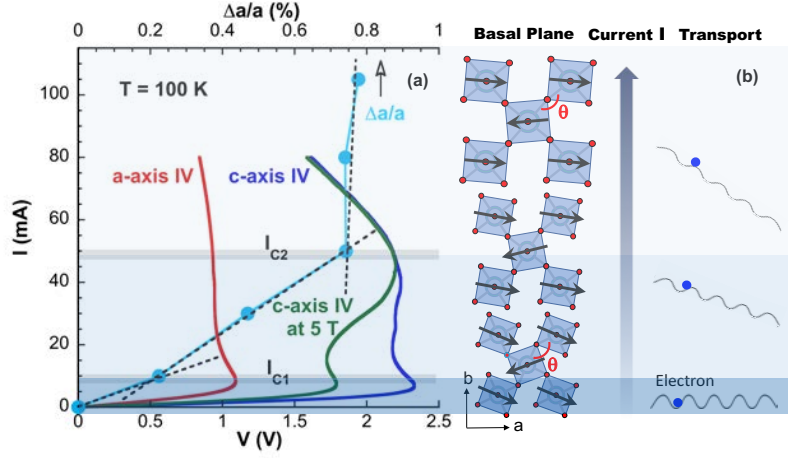

*Fig. 17. Sr$_2$IrO$_4$: (a)* I-V curves (red, blue, green) for a- and c-axis at T = 100 K and applied magnetic field of 0 or 5 Tesla along the c-axis. Light blue data (upper horizontal axis) show current-controlled a-axis expansion Δa/a at T = 100 K. Dashed lines are guides to the eye. Note slope changes of Δa/a occur at the two turning points of the I-V curves at I$_{C1}$ and I$_{C2}$, respectively. *(b)* Diagrams (not to scale) illustrate the expanding lattice, increasing Ir1-O2-Ir1 bond angle θ (red) and decreasing magnetic canting (black arrows) with increasing I. The reduced lattice distortions lead to enhanced electron mobility (see schematic) [6].

Nevertheless, the current-controlled *a*-axis expansion Δ*a*/*a* (upper horizontal axis in **Fig. 17a**) closely tracks the I-V curves with non-linear changes at two critical currents I$_{C1}$ (= 10 mA) and I$_{C2}$ (= 45 mA), respectively [6]. The slope changes in Δ*a*/*a* indicate successively more rapid expansions of the *a*-axis at I$_{C1}$ and I$_{C2}$, and each of them is accompanied by a more significant increase in the Ir-O-Ir bond angle θ, which in turn improves electron hopping (**Fig. 17b**). The *a*-axis expansion Δ*a*/*a* seems to saturate as the current further increases above I$_{C2}$ = 45 mA, suggesting that the lattice parameters cannot further expand at I > I$_{C2}$. This explains why a magnetic field H reduces V considerably only between I$_{C1}$ and I$_{C2}$ but shows no visible effect above I$_{C2}$ (green curve in **Fig. 17a**), where the saturation of Δ*a*/*a* corresponds to θ approaching 180°, which prevents further increases.

The correlation between the *a*-axis expansion (**Fig.12**), the magnetic canting (**Figs.9** and **16**) and the I-V curves (**Fig.15**) highlights a crucial role of the current-controlled basal-plane expansion that dictates the quantum states. In essence, this is because the $IrO_6$-octahedra and the canted moment are locked together, thus rigidly rotate together (**Figs. 9** and **10**) due to strong SOI; applied electrical current effectively engages with the lattice and expands the basal plane by increasing θ, which in turn reduces the magnetic canting and the AFM transition $T_N$ and enhances the electron mobility, as illustrated in **Fig.17**.

It is important to be pointed out that current-controlled phenomena are essentially absent in the spin-orbit-coupled $Sr_3Ir_2O_7$ (**Fig.3**) [35], a sister compound of $Sr_2IrO_4$, in part because of its collinear magnetic structure, which is discussed in *Section IV*. Recent studies reveal that current-controlled phenomena exist in a range of Mott systems having magnetic and/or structural distortions.

### C.     *Model System Two: Structurally-Driven Mott Insulator $Ca_2RuO_4$*

Structurally-driven Mott insulator $Ca_2RuO_4$ [24, 74] is another model system [5,7,8]. As discussed in *Section II.A*, this material exhibits a metal-insulator transition at $T_{MI}$ = 357 K driven by a violent structural transition [25]. This transition is accompanied by a severe rotation and tilting of $RuO_6$ octahedra, resulting in a considerably enhanced orthorhombicity below $T_{MI}$, and lifting the $t_{2g}$ orbital ($d_{xy}$, $d_{yz}$, $d_{zx}$) degeneracy [25, 26, 75-84]. An AFM state is only stabilized at a much lower temperature, $T_N$ =110 K [24, 74] by a further rotation and tilting of $RuO_6$ octahedra. It is now well-recognized that physical properties are dictated by the structural distortions and the populations of $t_{2g}$ orbitals, particularly, the $d_{xy}$ orbital [26, 70, 74-87].

We choose 3% Mn doped $Ca_2RuO_4$ or $Ca_2Ru_{0.97}Mn_{0.03}O_4$ for this investigation because the dilute Mn doping weakens the violent first-order structural phase transition at 357 K but preserves the underlying structural and physical properties of Ca$_2$RuO$_4$. This way, the single crystals are more robust to sustain thermal cycling necessary for thorough measurements [70, 87]. Note that a key effect of the slight Mn or other 3d ion doping on Ca$_2$RuO$_4$ is the weakening of the severe orthorhombicity while preserving the low-temperature orthorhombic symmetry (*Pbca*). Since the doping level is very low, the carrier concentration remains essentially unchanged [70, 87].

As discussed in *Section III. B*, the key element here is a novel coupling between small applied electrical current and the lattice that, in an extraordinary fashion, reduces the orthorhombic distortion, the octahedral rotation and tilt. These lattice changes in turn readily suppress the native AFM state and subsequently induce a nonequilibrium orbital state. A phase diagram generated based on the data illustrates a narrow critical regime near a small current density of 0.15 A/cm$^2$ that separates the native, diminishing AFM state and the emergent, nonequilibrium orbital state. It is particularly significant that a direct correlation between the current-reduced orthorhombicity and electrical resistivity is established via simultaneous measurements of both neutron diffraction and electrical resistivity. Notably, no current-induced diamagnetism, which is reported to exist in $Ca_2RuO_4$ [5], is observed in both doped and undoped $Ca_2RuO_4$ in this study.

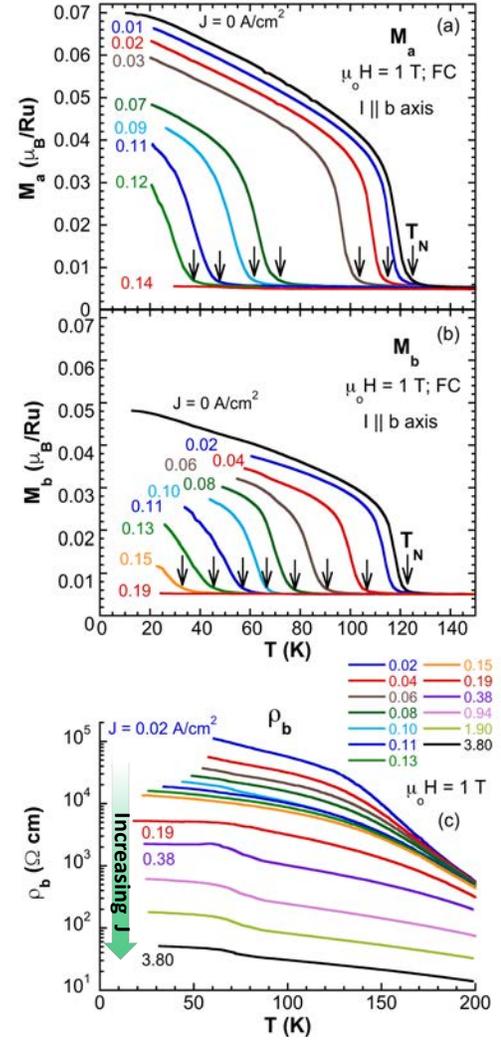

*Fig.18. Current-driven magnetic and transport properties of Ca$_2$Ru$_{0.97}$Mn$_{0.03}$O$_4$: The temperature dependence at various J applied along the b axis of (a) the a-axis magnetization M$_a$, (b) the b-axis magnetization M$_b$ and (c) the b-axis resistivity $\rho_b$. The magnetic field is at 1 T [8].*

In the following, we first discuss current-induced changes in the magnetization, electrical resistivity and then the lattice modifications and the correlation between them.

### 1. Current-controlled magnetization and electrical resistivity

As shown in **Figs.18a-18b**, the *a*- and *b*-axis magnetization, $M_a$ and $M_b$, changes drastically with electrical current applied along the *b* axis. The AFM transition $T_N$ drops rapidly from 125 K at current density J = 0 A/cm² to 29 K at J = 0.15 A/cm² in $M_b$, and completely disappears at J > $J_C$ ~ 0.15 A/cm² [8] (Note that 0.15 A/cm² is a remarkably small current density!) The vanishing AFM state is accompanied by a drastic decrease in the *b*-axis resistivity, $\rho_b$, by up to four orders of magnitude (**Fig.18c**). Note

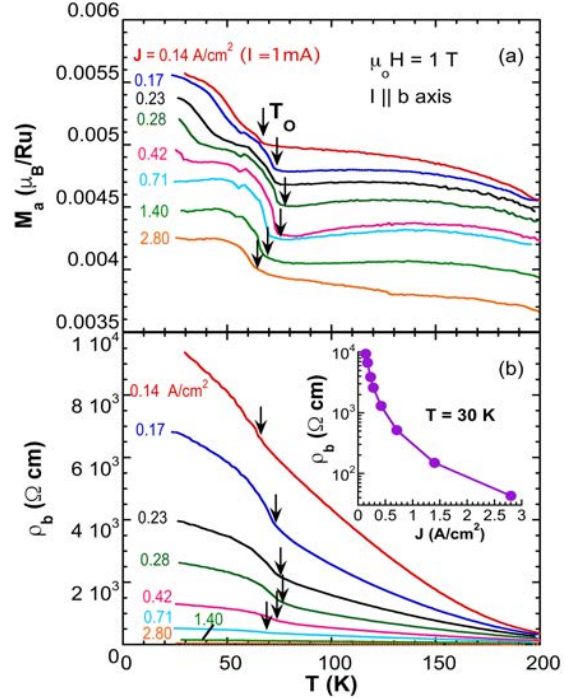

***Fig.19. Current-induced ordered state at J ≥ 0.14 A/cm²:*** *The temperature dependence at various J applied along the b axis of **(a)** $M_a$ at 1 Tesla and **(b)** $\rho_b$ for $Ca_2Ru_{0.97}Mn_{0.03}O_4$; **Inset**: $\rho_b$ at 30 K as a function of J [8].*

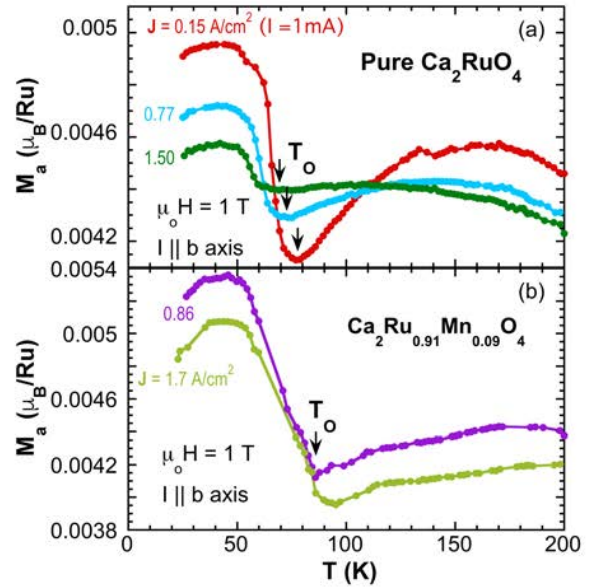

***Fig.20. Pure $Ca_2RuO_4$ and $Ca_2Ru_{0.91}Mn_{0.09}O_4$ for comparison: (a)*** *The a-axis magnetization $M_a$ at 1 Tesla at a few representative J for **(a)** pure $Ca_2RuO_4$ and **(b)** $Ca_2Ru_{0.91}Mn_{0.09}O_4$ [8].*

that the resistivity and magnetization are simultaneously measured.

A few points are particularly remarkable and deserve additional discussion.

### *(a) Emergent nonequilibrium orbital state*

Upon the vanishing of the AFM transition $T_N$, a distinct phase characterized by $T_O$ emerges below 80 K (see **Fig.19**). The phase transition $T_O$ rises initially, peaks at $J = 0.28$ A/cm$^2$ before decreasing with increasing J (**Fig.19a**). The *b*-axis resistivity $\rho_b$ intimately tracks the magnetization $M_a$ (**Fig.19b**). The *concurrent change* in both $\rho_b$ and $M_a$ at $T_O$ indicates a strong correlation between the transport and magnetic properties in the emergent state, which sharply contrasts the native state in which $T_N$ happens nearly 250 K below $T_{MI}$ [24, 25, 75], implying that the nature of the current-induced state is distinctly different from that of the native state. This is consistent with the fact that it emerges only when the equilibrium AFM state completely vanishes.

Above To, the magnetization shows a history-dependence (not shown), which is most likely associated with a metastable state. It is already established that the current-reduced structural distortions effectively diminish the AFM at $J \geq 0.15$ A/cm$^2$ (**Fig.18**), thus favoring a competing FM state [77, 80]. Further increasing J inevitably enhances the competition between the FM and AFM interactions but in a metastable manner, leading the history-dependent behavior. The neutron diffraction rules out any current-induced inhomogeneous effect [8].

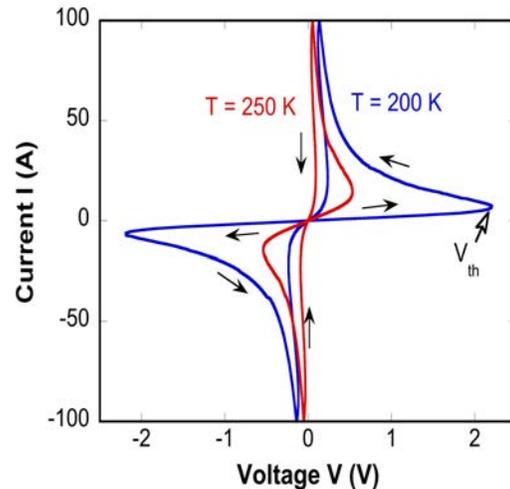

***Fig.21.*** *I-V characteristics for $Ca_2Ru_{0.97}Mn_{0.03}O_4$ at T = 200 and 250 K. Note that the I-V curves are taken above $T_O$ (~ 80 K).*

### *(b) No diamagnetism discerned*

It is stressed that the emergent state at J > $J_C$ is also observed in undoped Ca$_2$RuO **(Fig. 20a)**, and 9% Mn doped Ca$_2$RuO$_4$ **(Fig.20b)**. This indicates that the new state is a robust response of Ca$_2$RuO$_4$ to applied current, independent of Mn doping. Importantly, the data in **Figs.19-20** show no sign of the diamagnetism prominently reported in *Science 358, 1084 (2017)* or Ref. [5]. A controlled study on other magnetic materials [8] further buttresses the conclusion: No current-induced diamagnetism in both undoped and doped Ca$_2$RuO$_4$ [8]. Such a major discrepancy between results in Refs. [5] and [8] deserves more investigations.

*(c) Quantum switching effect*

As in Sr$_2$IrO$_4$, a sharp, unusual switching effect is also a pronounced feature in Ca$_2$RuO$_4$, as shown in **Fig.21** (Note that the data is taken from *Ca$_2$Ru$_{0.97}$Mn$_{0.03}$O$_4$* because it can sustain more thermal cycles than undoped Ca$_2$RuO$_4$). The mechanism of the switching effect is likely similar to that operating in Sr$_2$IrO$_4$, i.e., the current-induced lattice expansion drives the I-V characteristics, as discussed in *Section III. B*.

*2. Current-controlled structural properties*

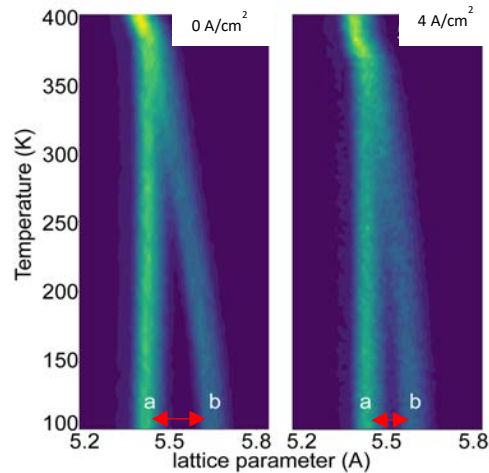

*Fig.22. The neutron diffraction and current-reduced orthorhombicity in Ca$_2$Ru$_{0.97}$Mn$_{0.03}$O$_4$:* Two representative contour plots for the temperature dependence of the lattice parameters a, and b axis at current density J = 0 and 4 A/cm$^2$ applied in the basal plane. Note the diminishing orthorhombicity with increasing J marked by the red arrows [8].

The structural properties as functions of applied current and temperature are thoroughly studied via neutron diffraction. A major effect is the diminishing orthorhombicity with increasing J, as shown in **Fig. 22.** Indeed, the orthorhombicity, defined by (*b-a*)/[(*a+b*)/2], readily reduces with increasing J -- from 4.4% at J = 0 A/cm$^2$ to 1.2% at J

= 30 A/cm$^2$ (**Figs.23a-23b**). At the same time, the *c* axis expands by up to 2.4% at J = 30 A/cm$^2$ (**Fig.23c**). It is also crucial that the bond angle Ru-O-Ru, which defines the rotation of RuO$_6$ octahedra, increases by up to two degrees at J = 18 A/cm$^2$, giving rise to much less distorted RuO$_6$ octahedra (**Figs.23d**). Furthermore, the bond angle O-Ru-O decreases from 91º to 90.2º at J = 5 A/cm$^2$, close to the ideal O-Ru-O bond angle of 90º (**Figs.23d**). The significantly relaxed crystal structure explains the transport and magnetic data in **Figs.18-21**. This strong, direct association is further illustrated below.

### 3. Correlations between current-controlled structural and physical properties

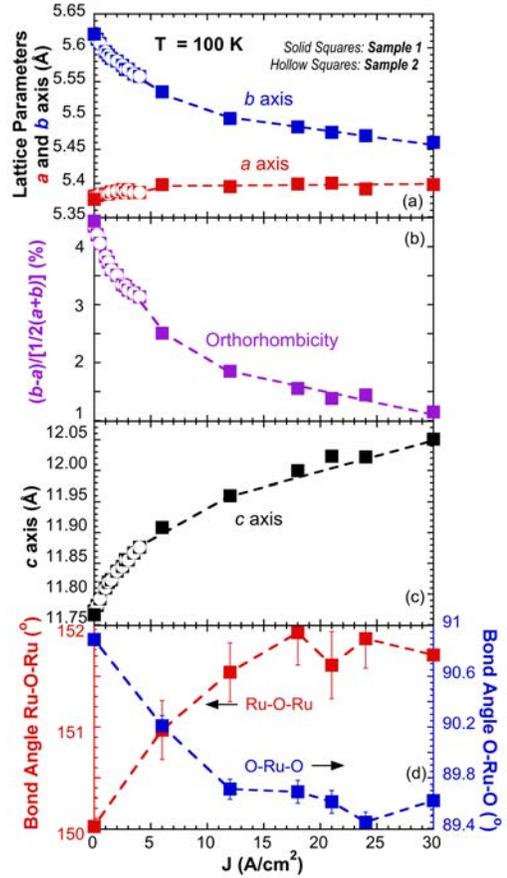

*Fig.23. The neutron diffraction and current-driven lattice changes in Ca$_2$Ru$_{0.97}$Mn$_{0.03}$O$_4$ The current density J dependence at 100 K of (a) the a and b axis, (b) the orthorhombicity, (c) the c axis and (d) the bond angles Ru-O-Ru (red, left scale) and O-Ru-O (blue, right scale)[8].*

The simultaneous measurements of neutron diffraction and electrical resistivity highlight a direct link between the current-reduced orthorhombicity and resistivity (see **Fig.24**). The orthorhombicity as functions of temperature and current density ranging from 0 to 4 A/cm$^2$ in **Fig. 24a** shows that the orthorhombic distortion rapidly reduces with current density J (< 1 A/cm$^2$). At the same time, the resistivity almost perfectly tracks the orthorhombicity, as shown in **Fig.24b**: *The contour-plot comparison of **Figs.24a-24b** compellingly establishes an explicit correlation between the current-driven lattice and transport properties*. Indeed, the improved structure significantly improves the $t_{2g}$ orbital occupancies for better electron hopping, as also evidenced in **Figs.18-19**.

It also deserves to be stressed that the structural transition $T_{MI}$, which is defined by the blue area in **Fig.24a**, hardly shifts with the current density J. This is also true for the data in **Fig.22**. These data along with the data in **Figs.20**, discussed earlier, confirm that Joule heating effect is inconsequential in these studies [6,8].

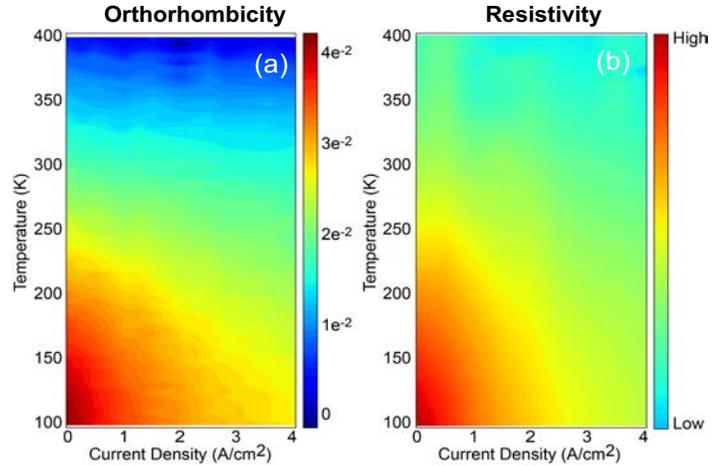

*Fig.24. Direct correlation between the orthorhombicity and the electrical resistivity of $Ca_2Ru_{0.97}Mn_{0.03}O_4$: The temperature-current-density contour plots for (a) the orthorhombicity and (b) electrical resistivity [8].*

## 4. Phase diagram and nonequilibrium orbital state

A temperature-current-density phase diagram generated based on the data presented above summarizes the current-controlled phenomena in $Ca_2RuO_4$: By reducing the structural distortions and changing $t_{2g}$ orbital occupancies, the applied current effectively destabilizes the insulating and AFM state and, at a narrow critical regime of current density $J_C$, precipitates the nonequilibrium orbital state, as shown in **Fig. 25**.

In ambient conditions, the tetravalent $Ru^{4+}$ ion with four *4d* electrons ion provides 2 holes in the $t_{2g}$ orbitals (with empty $e_g$ orbitals). A ½-hole is transferred to the oxygen [1], and the remaining 3/2 hole is equally split in a 1:1 ratio between the $d_{xy}$ orbital and the manifold of $d_{xz}/d_{xz}$ orbitals at high temperatures or in the metallic state at $T > T_{MI}$. At $T < T_{MI}$, the first-order transition $T_{MI}$ = 357 K enhances the orthorhombicity and other distortions including the rotation, tilting and flattening of octahedron $RuO_6$. These changes facilitate a transfer of more holes from $d_{xy}$ to $d_{xz}/d_{yz}$, or a 1:2 ratio of hole occupancies in $d_{xy}$ vs $d_{xz}/d_{yz}$ [1]. The insulating state below $T_{MI}$ thus has each orbital at exactly 3/4 electron filling. This contrasts the metallic state above $T_{MI}$ which has a nearly half-filled $d_{xy}$ orbital and unequal filling, with a nearly filled $d_{xz}/d_{yz}$ manifold.

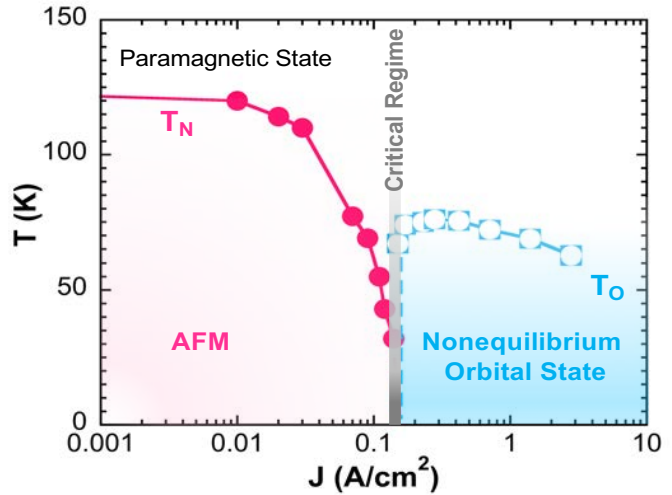

*Fig.25. The T-J phase diagram illustrates that the applied current drives the system from the native AFM state (purple) through the critical regime near 0.15 A/cm² (gray) to the current-induced, nonequilibrium orbital state (blue) [8].*

The half-filled orbital $d_{xy}$ is the key to the metallic state. The applied current helps stabilize the existence of the half-filled orbital $d_{xy}$ as temperature decreases by minimizing structural distortions (orthorhombicity, octahedral rotation and tilt). These current-induced lattice changes also explain the vanishing native AFM state with increasing J because it delicately depends on a combination of rotation, tilt and flattening of $RuO_6$ octahedra [77-84], all of which are significantly weakened by applied current, as shown in **Figs.22-24**.

While the understanding of the nonequilibrium orbital state at $J > J_C$ (**Fig.25**) is yet to be fully established, it is clear that at the heart of the current-controlled phenomena are the critical lattice modifications via current-driven nonequilibrium orbital populations.

### IV.    Challenges and Outlook

Mounting experimental evidence indicates that current controlled phenomena are widespread and present in a range of high-Z materials including both oxides and chalcogenides having an (anti)ferromagnetic and insulating state [88]. Recent studies point out a few empirical trends.

First of all, current-controlled materials must have comparable SOI and U and a ground state that is both (anti)ferromagnetic and insulating. It is apparent that *4d/5d* materials provide an energy setting more desirable for current control of quantum states because the *4d/5d* electronic wave functions are more extended, the *d*-band width is wider, and U and $J_H$ are smaller than those in the *3d* transition metal oxides (**Table 1**), as discussed in *Section II*. In these materials, the insulating and magnetic ground state is not enabled by large U but rather driven by subtle interactions which are assisted by SOI; thus, a small external stimulus such as electrical current could be sufficient to produce a large response, leading to phase transitions, as illustrated in the cases of $Sr_2IrO_4$ and $Ca_2RuO_4$.

Secondly, an effective current control of quantum states also requires the presence of a distorted crystal structure, oftentimes, canted moments and a strong magnetoelastic coupling. A distorted structure such as rotations and/or tilts of octahedra $MO_6$ generates room for applied current to relax correlated octahedral rotations and/or tilts and magnetic structures via a strong magnetoelastic coupling that locks magnetic moments with the lattice through SOI [58-60]. This point is demonstrated in $Sr_3Ir_2O_7$ [35] in which current-controlled behavior is essentially absent. This double-layered iridate is a sister compound of $Sr_2IrO_4$ and an AFM insulator with $T_N$=285 K and equally strong SOI (**Figs.3a** and **3b**) [4, 35]. However, the magnetic moments in $Sr_3Ir_2O_7$ are collinearly aligned along the *c* axis [89, 90] rather than canted within the basal plane where the octahedral rotation occurs, contrasting with that of $Sr_2IrO_4$. As a result, applied current cannot exert a sufficient effect because the collinearly aligned magnetic moments along the *c* axis is not strongly coupled with the $IrO_6$ rotation. Therefore, $Sr_3Ir_2O_7$ hardly shows current-controlled behavior observed in $Sr_2IrO_4$.

In short, the search of current-controlled materials should focus on AFM Mott insulators with strong structural and/or magnetic distortions in high-Z materials *where the role of SOI is significant and electron orbitals are extended* – SOI lock magnetic moments to the lattice and the extended orbitals facilitate a strong coupling of current and electron orbitals. An effective current-control of quantum states is not anticipated in low-Z materials such as *3d* materials because of the lack of the key elements.

While current-controlled phenomena and materials pose tantalizing prospects for unique functional materials and devices, a better understanding of them must be established. Clearly, theoretical efforts are urgently needed to help gain more insights into the physics of the nonequilibrium phenomena in correlated and spin-orbit-coupled materials.

Nevertheless, the empirical trends observed in recent years help pose a series of intriguing questions that may provide the impetus for advancing our understanding of this class of materials:

- *Current-driven phenomena are essentially nonequilibrium phenomena, which are both exciting and intellectually challenging; how can we more comprehensively tackle the challenge both theoretically and experimentally?*
- *In particular, how can we adequately describe the coupling of current and the lattice and/or magnetic moments in this class of materials? This is among key issues for this research topic and needs to be addressed urgently.*
- *Can an applied current increase the mixing of the $J_{eff} = 1/2$ and $J_{eff} = 3/2$ states and be responsible for the dramatic current-induced changes in the iridates?*
- *While the non-Ohmic I-V characteristics are closely associated with the current-induced lattice expansion experimentally, how can we adequately describe changes in band structures fundamentally responsible for them?*

- *How can we more effectively current-control structural, magnetic and transport properties?*

- *More generally, can we establish a set of criteria to identify and improve these materials?*

- *What are potential applications of this class of materials?*

- *What unique devices can we propose or develop using the novel current-controlled properties?*

It is clear that current control of quantum states opens new avenues for better understanding the fundamental physics of spin-orbit-coupled matter. Equally importantly, it also provides a new paradigm for the development of an entire class of current-controlled materials to underpin functional devices otherwise unavailable.

**Endnote** Upon the completion of this *Topical Review*, a newly posted *arXiv* manuscript, which is authored by the same research group that reported the diamagnetism in *Science 358, 1084 (2017)* or Ref. [5], attributes the reported diamagnetism to *spurious behavior* rather than intrinsic response to the applied current [91]. More recently, the authors of *Science 358, 1084 (2017)* and *Physical Review Letters 122, 196602 (2019)* (on diamagnetism in $Ca_3(Ru_{1-x}Ti_x)_2O_7$) retract the two papers [92, 93].

**Acknowledgments** The author is indebted to Mr. Hengdi Zhao and Dr. Bing Hu for useful discussions when writing this *Topical Review*. This work is supported by the US National Science Foundation via grant DMR-1903888.